\documentclass[twocolumn,superscriptaddress,prB]{revtex4-1}
\usepackage{graphicx}
\usepackage[hidelinks]{hyperref}
\pdfoutput=1

\begin{document}
\title{Band gap control in phosphorene/BN structures from first principles calculations}

\author{Lukas Eugen \surname{Marsoner Steinkasserer}}
\email{marsoner@zedat.fu-berlin.de}
\affiliation{Institut f\"{u}r Chemie und Biochemie, Freie Universit\"{a}t Berlin, Takustra\ss e 3, D-14195 Berlin, Germany}

\author{Simon Suhr}
\affiliation{Institut f\"{u}r Chemie und Biochemie, Freie Universit\"{a}t Berlin, Takustra\ss e 3, D-14195 Berlin, Germany}

\author{Beate Paulus}
\affiliation{Institut f\"{u}r Chemie und Biochemie, Freie Universit\"{a}t Berlin, Takustra\ss e 3, D-14195 Berlin, Germany}

\begin{abstract}

Using both DFT as well as $G_0W_0$ calculations, we investigate static and dynamic effects on the phosphorene band gap upon deposition and encapsulation on/in BN multilayers. We demonstrate how competing long- and short-range effects cause the phosphorene band gap to increase at low $\rm{P}-\rm{BN}$ interlayer spacings, while the band gap is found to drop below that of isolated phosphorene in the BN/P bilayer at intermediate distances around 4 \AA.
Subsequent stacking of BN layers, i.e. BN/BN/P and BN/BN/BN/P is found to have a negligible effect at the DFT level while at the $G_0W_0$ increased screening lowers the band gap as compared to the BN/P bilayer. 
Encapsulation between two BN layer is found to increase the phosphorene band gap by a value approximately twice that observed when going from freestanding phosphorene to BN/P. 
We further investigate the use of the GLLB-SC functional as a starting point for $G_0W_0$ calculations showing it to, in the case of phosphorene, yield results close to those obtained from $GW_0@\textnormal{PBE}$ calculations.

\end{abstract}

\date{July 27, 2016}
\maketitle

\section{Introduction}

In 2014 a new type of material consisting of a black-phosphorus monolayer (phosphorene) was first synthesized  \cite{li2014black,liu2014phosphorene,xia2014rediscovering,qiao2014high}. Phosphorene has quickly become the subject of considerable research interest due to its attractive optical gap between 1.3 eV  \cite{wang2015highly}, 1.45 eV  \cite{liu2014phosphorene}, 1.75 eV  \cite{yang2015optical} and 1.73 eV \cite{li2016direct} as well as high charge-carrier mobility  \cite{kou2015phosphorene}. The material is further made appealing by the anisotropy of its transport properties, with electrons and holes possessing notably different effective masses along the phosphorene armchair and zig-zag direction.

While being extremely attractive as an object of study, pristine phosphorene unfortunately suffers from rapid degradation by oxygen under ambient conditions  \cite{island2015environmental}. A natural remedy for this problem consists in protecting the phosphorene layer by capping or encapsulating it with/in more environmentally stable materials.
 BN has been proposed as the natural candidate for this application and BN/phosphorene heterostructures have been studied in a number of recent theoretical works at the DFT level \cite{cai2015electronic,hu2015anisotropic,you2015prediction, constantinescu2016multipurpose}.

It is well known though from other 2D-materials that adsorption, even on materials possessing low dielectric constants such as BN, can have a significant impact on their electronic properties  \cite{jiang2013giant}. As these effects are often attributable to long-range screening effects not accounted for within DFT, \emph{many-body} electronic structure methods such as \emph{GW} are necessary to study them from a theoretical point of view.

Herein, we consider in detail the effects of the interaction of phosphorene with BN on the phosphorene electronic structure employing recently developed methods for the accurate computation of electron-correlation effects within 2D-materials  \cite{rasmussen2015efficient}. The effects of screening are investigated for different numbers of layers of BN on both sides of phosphorene with particular emphasis being placed on the effects of the P/BN interlayer spacing as well as different P/BN stacking sequences.

\section{Computational Details}
The lattice parameters for phosphorene and BN were obtained using the PBE0 \cite{adamo1999toward} functional combined with Grimme's \emph{D2} dispersion correction \cite{grimme2006semiempirical}. We chose this combination of methods as it was shown by Sansone et al.  \cite{sansone2015towards} to yield lattice parameters very close to those obtained from high-level \emph{ab-initio} calculations. Relaxation of the monolayer was performed using the CRYSTAL14 program  \cite{dovesi2014crystal14,crystal14man} together with a POB-triple-$\zeta$ basis set as described by Peintinger et al. \cite{peintinger2013consistent}. 

\begin{figure}[h]
\centering
  \includegraphics[width=0.48\textwidth,angle=0]{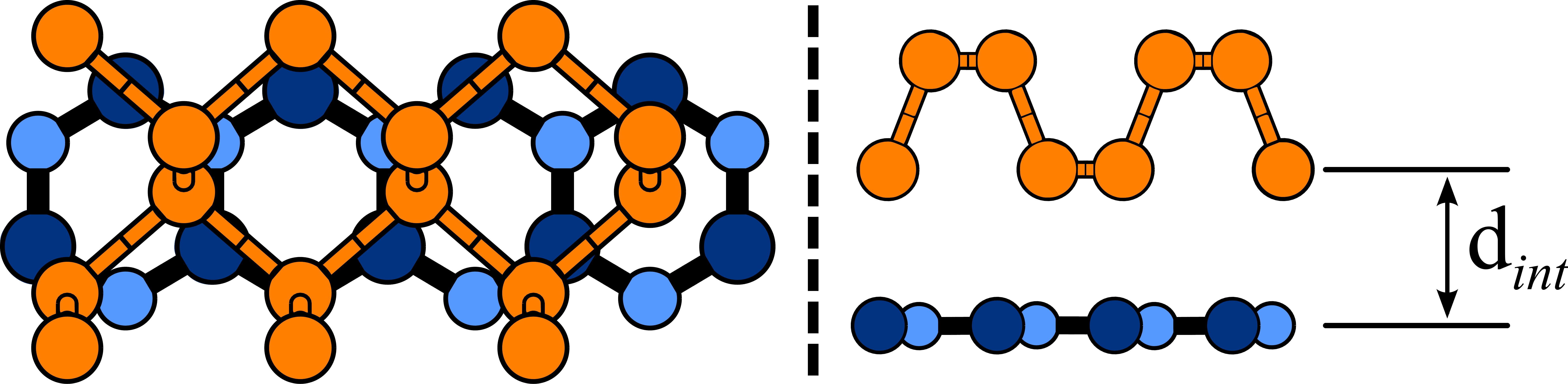}
  \caption{The left-hand figure shows a top-on view of the phosphorene/BN bilayer supercell used for all multilayer calculations throughout this work, while the right-hand site shows a site-on view of the same structure where the definition of the interlayer spacing has also been indicated.}
\label{bilayer_positions}
\end{figure}

Heterostructures were modeled via a unit cell consisting of a $1\times3$ phosphorene supercell and a $1\times4$ supercell of the primitive orthorhombic cell of BN. The resulting cell is shown in figure \ref{bilayer_positions}. The maximum resulting strain within the BN layers is $\approx 4.7$ \%. Note that we have tested the robustness of our conclusions against a consistent change in the lattice constant of the phosphorene monolayer and the P/BN heterostructure and found the effect to be that of a rigid shift of the band gap with only a small ($\approx 0.03$ eV) difference in the change of the band gap upon formation of the bilayer.

Plane-wave DFT calculations were performed using the GPAW code  \cite{bahn2002object,mortensen2005real,enkovaara2010electronic, PhysRevB.87.235132} and employed an energy cutoff of \mbox{500 eV} using a vacuum spacing of 10 \AA. Electron correlation effects on the band gap were taken into account using the $G_0W_0$ method \cite{hedin1965new} as implemented in GPAW including recently developed methods accounting for the dielectric functions long-wavelength behavior in 2D systems  \cite{rasmussen2015efficient, huser2013dielectric}. \emph{GW} Calculations were performed using a $8\times12$ k-grid while the $GW$-self energy was computed at three cutoff values up to 110 eV (95 eV, 102 eV, 110 eV) and subsequently extrapolated to infinity  \cite{schindlmayr2013analytic, klimevs2014predictive}. 

For multilayer calculations a $8\times4$ k-grid was used. In all cases the frequency dependence was represented on a non-linear grid from 0 eV to the energy of the highest transition included in the basis set. The grid-spacing was gradually increased starting from 0.15 eV and reaching 0.3 eV at 2 eV. Lastly, as pointed out by Rasmussen et al. the efficiency of the analytical correction around $\mathbf{q}=0$  depends on the length of the unit cell along the non-periodic direction. This causes a slowing of the k-grid convergence, when the vacuum spacing is increased. Though this effect is small at the range of vacuum-spacings considered in this work (an increase of 6 \AA\ in the length of the unit cell along the non-periodic direction resulting in an increase in the band gap of $\approx 0.04$ eV) we chose to correct for it by performing calculations on isolated phosphorene monolayers at their relaxed geometry within the same supercell used for $GW$ calculations on P/BN multilayers. The multilayer gap was then corrected by the difference between the phosphorene monolayer in the supercell and that obtained for the phosphorene unit cell. This procedure was applied to all multilayer structures considered herein.  It is worth noting that this correction is not exact and might lead to an under- or overestimation of the band gap. While this does slightly alter the observed band gap behavior it does not influence our general conclusions. As an example the difference between the BN/BN/BN/P and BN/BN/P band gap is reduced from -0.05 eV to -0.02 eV by excluding the correction at the $G_0W_0@\textnormal{PBE}$ level.

Herein, we further present a possible low-cost alternative to the iteration of the \emph{GW} equations or the use of computationally demanding hybrid-DFT functionals as starting points for $G_0W_0$ calculations. Our proposed method consists in the use of the GLLB-SC functional  \cite{kuisma2010kohn} which provides a low-cost approximation to the EXX-OEP potential. We show how this allows to obtain close $GW_0$ quality results at the cost of $G_0W_0@\textnormal{GGA}$. The GLLB-SC functional further allows for the calculation of the quasiparticle gap ($E_g^{QP}$), i.e. the difference of the ionization potential and electron affinity, via the sum of the Kohn-Sham gap and the derivative discontinuity \cite{perdew1982density,kuisma2010kohn}. $E_g^{QP}$ values resulting from GLLB-SC$+\Delta_{xc}$ have been shown to yield results in close agreement with experimental results  \cite{castelli2012computational}.

Though the accuracy of band gaps is clearly improved by the inclusion of $\Delta_{xc}$ and the result of any perturbative method (such as $G_0W_0$) should benefit from an improved starting point, it is unclear as to whether $G_0W_0@\textnormal{GLLB-SC}+\Delta_{xc}$ should be considered as more accurate than results obtained from a pure Kohn-Sham starting point, i.e. $G_0W_0@\textnormal{GLLB-SC}$. This is because, as pointed out by Yan et al.  \cite{yan2012optical}, inclusion of the derivative discontinuity in the calculation of the dielectric constant at the RPA level leads to a systematic underestimation of the static screening and similar findings have been observed e.g. in the case of the HSE03 screened-hybrid functional  \cite{fuchs2007quasiparticle}.

Given this uncertainty regarding the correct computational method, herein we present results at the $G_0W_0@\textnormal{GLLB-SC}+\Delta_{xc}$, $G_0W_0@\textnormal{GLLB-SC}$ as well as $G_0W_0@\textnormal{PBE}$ level for all mono- and bilayer systems, while only $G_0W_0@\textnormal{GLLB-SC}$ and $G_0W_0@\textnormal{PBE}$ results are given in the tri- and quadrulayer case.

\begin{figure}[h]
\centering
  \includegraphics[width=0.48\textwidth,angle=0]{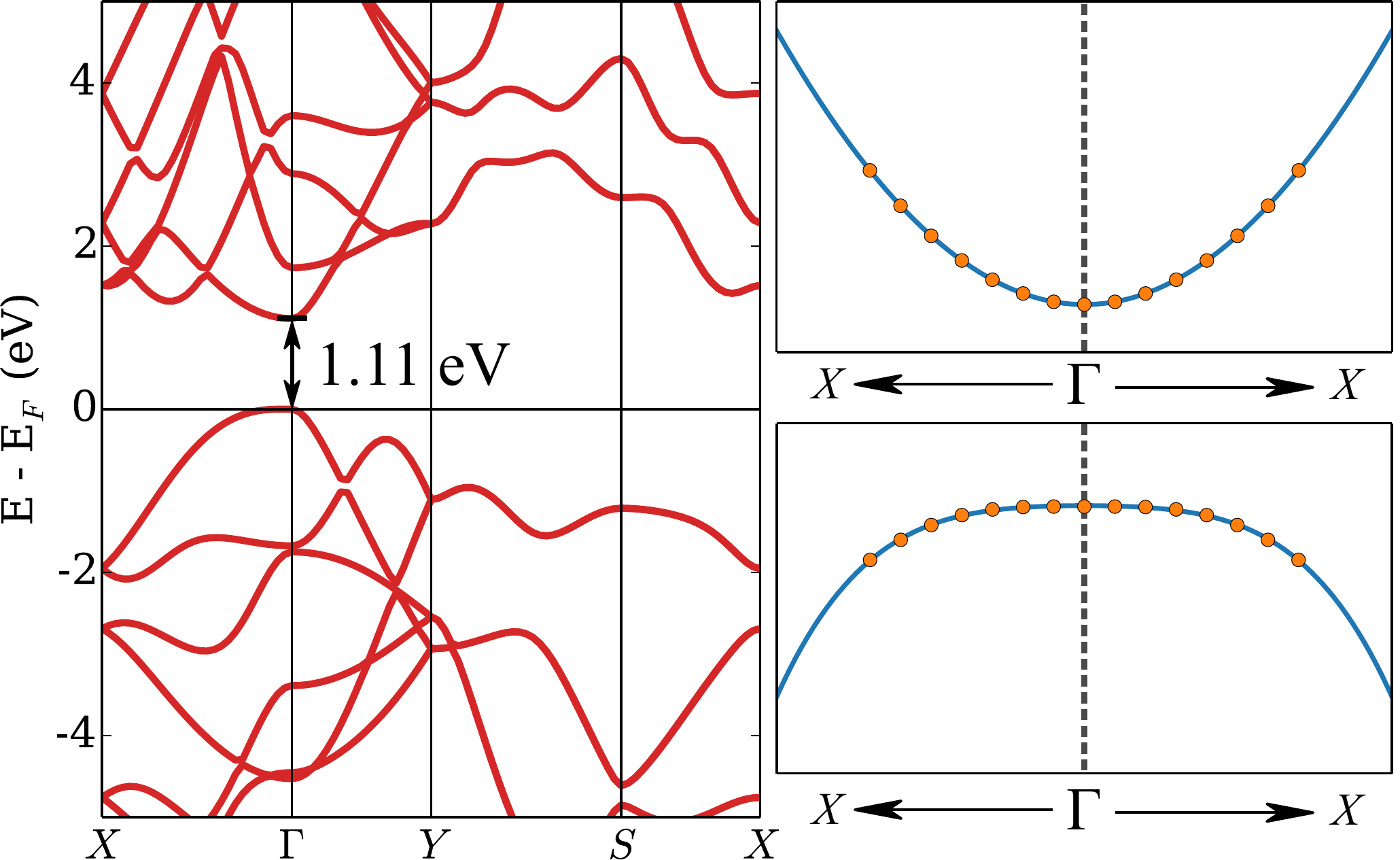}
  \caption{The left-hand figure shows the phosphorene band structure at the GLLB-SC level. On the right-hand site blowups of the region around the $\Gamma$-point show two examples of the fitting procedure used to obtain the electron/hole effective masses with the points corresponding to calculate band-structure values and the lines to the interpolating spline functions.}
\label{p_bands_gllb}
\end{figure}

\section{Results and Discussion}

\subsection*{Monolayer phosphorene}

We will begin our discussion by briefly discussing the electronic structure of the phosphorene monolayer. As mentioned previously, phosphorene displays anisotropic band dispersion around the $\Gamma$-point resulting in different effective masses for electrons/holes along the $\Gamma-X$ and $\Gamma-Y$ directions, respectively. The GLLB-SC band structure of phosphorene is shown in figure \ref{p_bands_gllb}.  

To quantify the difference in dispersion along $\Gamma-X$ and $\Gamma-Y$, we computed hole/electron effective masses along both high-symmetry directions by fitting the VBM/CBM around the $\Gamma$ point to splines of order four (using a fitting range of 0.15 a$_0^{-1}$) and calculating the second derivative at the $\Gamma$ point. 
For the $\Gamma-X$ direction, figure \ref{p_bands_gllb} shows a blowup of the phosphorene band structure around the $\Gamma$-point where the data points used for the fit to obtain the hole/electron effective mass as well as the fitted spline function are indicated.
Using this procedure, we obtained electron effective masses of 1.2 $m_e$ and 0.2 $m_e$ ($m_e$ being the electron rest mass) along the $\Gamma-X$ and $\Gamma-Y$ direction, respectively. The corresponding hole effective masses are 8.3 $m_e$ and 0.2 $m_e$. 

\begin{table}[h]
\centering
\begin{tabular}{@{}lcc@{}}
\hline\hline
Method & \begin{tabular}{@{}c@{}}
$a = 3.31$ \AA\\
$b = 4.52$ \AA
   \end{tabular}
&  \begin{tabular}{@{}c@{}}
$a=3.32$ \AA\\
$b=4.42$ \AA
   \end{tabular}\\ \cline{2-3} 
$G_0W_0@\textnormal{PBE}$                   &  1.89 eV          &   1.72 eV     \\
$G_0W_0@\textnormal{GLLB-SC}$               &  2.17 eV          &   2.01 eV     \\
$G_0W_0@\textnormal{GLLB-SC}+\Delta_{xc}$   &  2.77 eV          &   2.56 eV     \\
GLLB-SC$+\Delta_{xc}$                       &  1.62 eV          &   1.45 eV     \\ \hline\hline
\end{tabular}

 \caption{Band gap data for the phosphorene monolayer as obtained at different levels of theory. The two columns refer to calculation done at the PBE0-D2-relaxed lattice parameters of phosphorene ($a = 3.31$ \AA, $b = 4.52$ \AA) as well as the average between the PBE0-D2-relaxed phosphorene and BN lattice constants for the supercell-construction shown in figure \ref{bilayer_positions} ($a=3.32$ \AA, $b=4.42$ \AA). GLLB-SC $+\Delta_{xc}$ refers to the qusiparticle gap ($E_g^{QP}$) calculated using the GLLB-SC functional including the derivative discontinuity.}
\label{g0w0_p_mono}
\end{table}

These results differ only slightly from previous HSE06 data  \cite{hu2015anisotropic} the largest discrepancy being in the hole-mass along the $\Gamma-X$ direction for which Hu and Hong report a value of 7.43 $m_e$ in contrast to our value of 8.3 $m_e$ though this difference is unsurprising given the difference in method as well as flatness of the bands. 
While all other effective mass values are rather insensitive to the particular choice of fitting range, the VBM is highly non-harmonic along the $\Gamma-X$ direction leading to difficulties in obtaining an accurate value for the second derivative. This is indeed also reflected in the literature with phosphorene $\Gamma-X$ hole effective mass values ranging from the aforementioned 7.43 $m_e$  \cite{hu2015anisotropic} (HSE06) to 6.35 $m_e$  \cite{qiao2014high} (optB88-vdW), 4.92 $m_e$  \cite{peng2014strain} (PBE) and 1.61 $m_e$  \cite{choi2015linear} (PW91).

As a final stage in the analysis of the phosphorene monolayer we have calculated its $G_0W_0$ band gap using a number of different methods. Table \ref{g0w0_p_mono} provides a summary of all the data obtained. To demonstrate the applicability of our approach, we further performed a series of calculations using the same phosphorene lattice parameters as those used by Rasmussen et al. \cite{rasmussen2015efficient}. 

They obtain band gaps of \mbox{0.90 eV}, \mbox{2.03 eV} and \mbox{2.29 eV} at the PBE, $G_0W_0@\textnormal{PBE}$ and $GW_0@\textnormal{PBE}$ level respectively. Our values for PBE and $G_0W_0@\textnormal{PBE}$ match theirs and $G_0W_0@\textnormal{GLLB-SC}$ results in a band gap of \mbox{2.30 eV} thereby reproducing the results obtained from $GW_0@\textnormal{PBE}$ very closely. Lastly for the same structure we obtain a $G_0W_0@\textnormal{GLLB-SC}+\Delta_{xc}$ band gap of \mbox{2.94 eV}. The good agreement between $G_0W_0@\textnormal{GLLB-SC}$ and $GW_0$@PBE is especially encouraging given the aforementioned modest costs of the $G_0W_0@\textnormal{GLLB-SC}$ method as compared to $GW_0@\textnormal{PBE}$. 

We mention in passing the good results obtained from GLLB-SC$+\Delta_{xc}$ which gives a band gap only $\approx 0.26$ eV below the converged $G_0W_0@\textnormal{PBE}$ value. As the computational requirements for these calculations are on par with those of GGA calculations, they are a very good choice for cases in which \emph{GW} calculations are not feasible. Lastly, we note the decrease of $\approx 0.17$ eV in the phosphorene band gap as its lattice constant is modified to the average of the relaxed phosphorene and BN lattice constants. This also is reproduced well at the GLLB-SC$+\Delta_{xc}$ level.

\subsection*{Bilayers}

P/BN bilayers as schematically shown in figure \ref{bilayer_positions} are constructed following the lowest-energy relative orientation of phosphorene and BN determined by previous studies \cite{cai2015electronic,hu2015anisotropic}. Atomic positions were then relaxed at the PBE0-D2 level, while the supercell lattice constants were kept fixed at the relaxed phosphorene values ($a = 3.31$ \AA\ and $b = 4.52$ \AA).

\begin{figure}[h]
\centering
  \includegraphics[width=0.45\textwidth,angle=0]{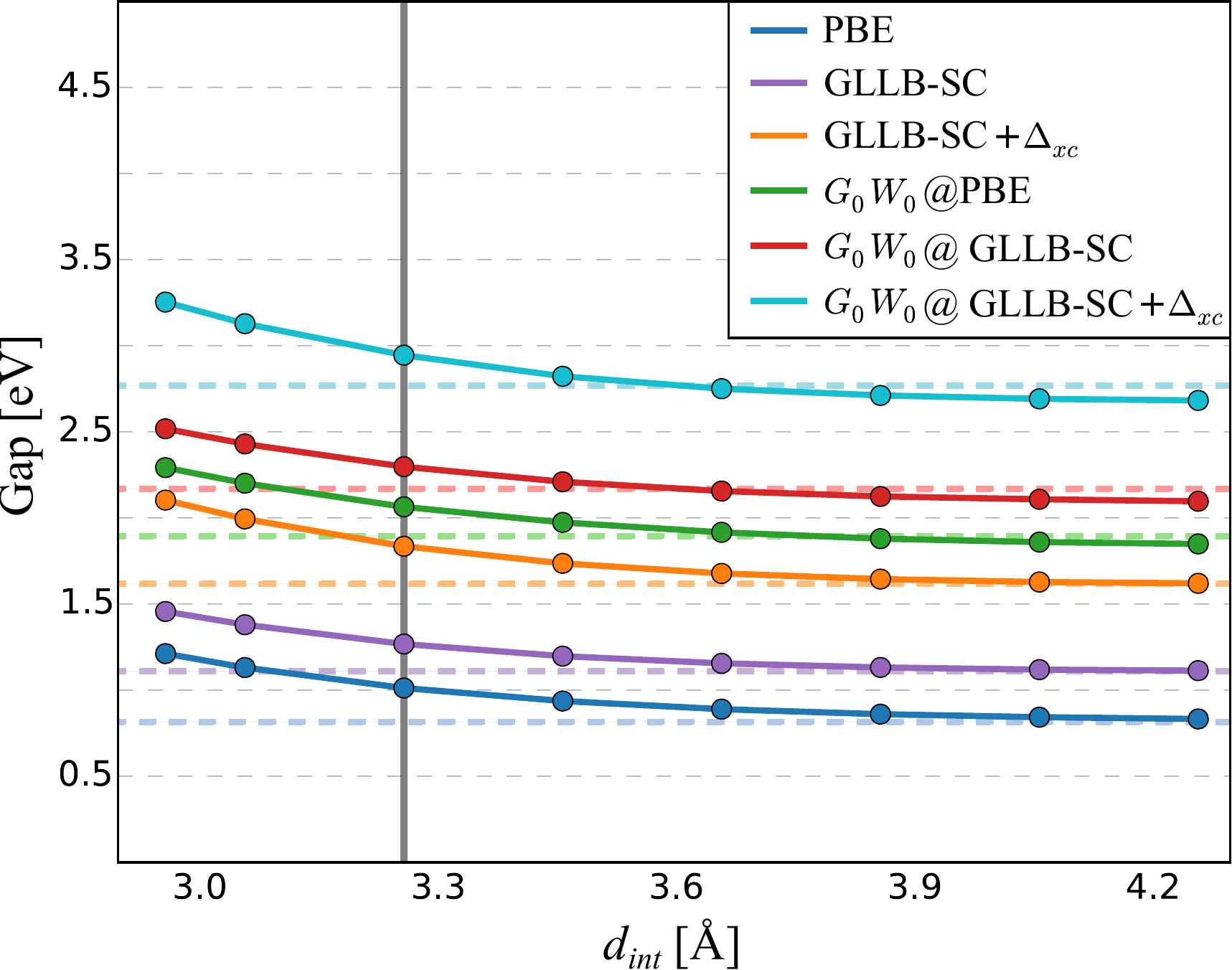}
  \caption{Phosphorene band gap as a function of the $\rm{P}-\rm{BN}$ interlayer spacing (see figure \ref{bilayer_positions}). Horizontal lines are drawn to indicate the value of the isolated phosphorene band gap at the different levels of theory while the vertical line indicates the relaxed PBE0-D2 interlayer spacing.}
\label{gap_dist_scan}
\end{figure}

\begin{table*}
\centering
\begin{tabular}{@{}lccccc@{}} 
\hline\hline
                      &    P     &  BN/P    &  BN/BN/P &  BN/BN/BN/P &  BN/P/BN \\ \cline{2-6} 
PBE                   &  0.82 eV &  1.01 eV &  1.03 eV &   1.03 eV   &  1.21 eV \\ 
GLLB-SC               &  1.11 eV &  1.27 eV &  1.28 eV &   1.28 eV   &  1.43 eV \\ 
GLLB-SC$+\Delta_{xc}$ &  1.62 eV &  1.83 eV &  1.85 eV &   1.85 eV   &  2.05 eV \\ 
$G_0W_0@\textnormal{PBE}$          &  1.89 eV &  2.06 eV &  2.04 eV &   1.99 eV   &  2.22 eV \\ 
$G_0W_0@\textnormal{GLLB-SC}$      &  2.17 eV &  2.30 eV &  2.26 eV &   2.22 eV   &  2.44 eV \\ \hline\hline
\end{tabular}
 \caption{Band gap data for the phosphorene/BN multilayer systems as obtained at different levels of theory. The $\rm{P}-\rm{BN}$ distance is almost identical (within 0.02 \AA) in all cases and equal to 3.2 \AA \ while $\rm{BN}-\rm{BN}$ distances are equal to 3.0 \AA. For comparison, data for the BN/P bilayer and phosphorene monolayer are also shown.}
\label{trilayer_gaps}
\end{table*}

Interlayer distances for vdW-bound systems are notoriously difficult to compute, with competing methods often providing strongly diverging values (see f.e. the case of black phosphorus  \cite{sansone2015towards}). The corresponding potential energy surfaces are also in general rather flat, which means systems can undergo significant changes in the interlayer spacing due to external pressure, making it important to obtain results for interlayer spacings other than the relaxed value for a given DFT functional. We therefore performed a series of calculations at different interlayer spacings by rigidly shifting phosphorene with respect to the BN layer.

Let us first consider the influence the $\rm{P}-\rm{BN}$ interlayer spacing on the phosphorene band gap at different levels of theory (see figure \ref{gap_dist_scan}). It is worth iterating at this point that, as already discussed in the literature \cite{cai2015electronic,hu2015anisotropic,you2015prediction}, the band gap for all structures considered in this work is clearly localized on the phosphorene layer as the BN-centered bands lie well outside of the band gap region. 

Let us now turn to a closer analysis of figure \ref{gap_dist_scan}. The first observation we make is the fact that the curves for all five methods are nearly parallel to one-another with an increase of the band gap with increased level of computation i.e. $\textnormal{PBE}\rightarrow\textnormal{GLLB-SC}\rightarrow\textnormal{GLLB-SC}+\Delta_{xc}\rightarrow G_0W_0\textnormal{@PBE}\rightarrow G_0W_0\textnormal{@GLLB-SC}\rightarrow G_0W_0\textnormal{@GLLB-SC}+\Delta_{xc}$ though this parallelism is not maintained fully throughout the curves. While both the PBE and GLLB-SC$+\Delta_{xc}$ band gap smoothly converge to their respective isolated phosphorene values, $G_0W_0$ calculations result in a drop below the value obtained of the phosphorene monolayer at interlayer spacings larger than $\approx 3.8$ \AA. This behavior can be understood if we consider the effects not accounted for within the DFT calculations. 
While the increase in band gap for low interlayer spacings is seen in all six curves and likely attributable to local electrostatic effects accounted for in both DFT as well as $GW$ calculations, only the latter correctly treat electron correlation. The dropping of the phosphorene band gap below its isolated monolayer value is therefore most likely due to slowly-decaying screening effects. Lastly we note that, at the relaxed PBE0-D2 interlayer distance, the band gap of the bilayer is increased with respect to that of freestanding phosphorene by about \mbox{0.2 eV} depending on the method (see table \ref{trilayer_gaps}).

\subsection*{Multilayers}

Given the strong dependence of the phosphorene band gap on the presence of a single BN layer, the question naturally arises as to how the addition of a second BN layer might influence the phosphorene's properties.
 Here we have considered two types of P/BN trilayer systems, i.e. the adsorbed (BN/BN/P) as well as encapsulated (BN/P/BN) system. In the former case the BN/BN stacking was chosen as AA' (i.e. AA stacking with B on top of N) in accordance with the experimentally most stable stacking \cite{constantinescu2013stacking} while in both cases the phosphorene layer was positioned on the BN layers as in the bilayer case shown in figure \ref{bilayer_positions}. 
Given the high computational demand of the calculations we have only performed $G_0W_0@\textnormal{GLLB-SC}$ and $G_0W_0@\textnormal{PBE}$ calculations, though results at the $G_0W_0@\textnormal{GLLB-SC}+\Delta_{xc}$ level are expected to be follow the same trends given the high degree of parallelism seen from the curves in figure \ref{gap_dist_scan}.

The resulting band gaps at different levels of theory are shown in table \ref{trilayer_gaps}. As expected from the discussion in the previous section, the addition of a second BN layer increases the band gap at the DFT-level for both functionals in the encapsulated as well as deposited case. The increase in the encapsulated system is though much larger than in the adsorbed one as in the former, the second BN layer is in direct contact with the phosphorene resulting in an additive effect from the two BN monolayers.
 In the deposited case however the large distance between phosphorene and the second BN layer causes the band gap to only increase slightly at the DFT level ($\approx 0.02$ eV). 
At the $G_0W_0$ level the effect is reversed as long-range screening effects more than compensate for this slight increase, pushing the band gap of the deposited trilayer slightly below that of the BN/P bilayer. 
In order to investigate whether this trend would continue if the number of BN layers was to be further increased we have performed calculations also on the BN/BN/BN/P (3$\times$BN/P) quadrulayer (see table \ref{trilayer_gaps}). Short-range DFT effects are already saturated at the trilayer and the 3$\times$BN/P DFT gap is identical to that of the BN/BN/P system. The $G_0W_0$ gap on the other hand is reduced by $\approx 0.05$ eV as compared to the BN/BN/P result, approaching the value of freestanding phosphorene. Unfortunately, it is not feasible to examin even larger BN layers to investigate the long-range decay of the screening.

\section{Conclusions}

We have studied in detail the interaction between phosphorene and BN, establishing a strong dependence of the phosphorene band gap on the P/BN interlayer spacing. In accordance with previous research \cite{cai2015electronic,hu2015anisotropic,you2015prediction} at the DFT-level, an increase in the band gap of phosphorene was observed upon adsorption on BN. On the other hand, long-range screening was found to significantly affect band gap in multilayered P/BN systems resulting in a lowering of the band gap. Given the variation in the band gap as well as the weakness of the interlayer binding one could envision employing this or similar stacked multilayers in the construction of pressure-sensitive devices in the future. We hope that this investigation will spark further research in counteracting static/dynamic effects on the band gap of vdW-multilayers with varying dielectric constants, thereby paving the way to future devices based on this type soft band gap control.

\begin{acknowledgments}
LEMS acknowledges the financial support by the the Studienstiftung des deutschen Volkes e.V., the grand provided by the Deutsche Forschungsgemeinschaft within the Priority Program (SPP) 1459 (Graphene) and the International Max Planck Research School "Complex Surfaces in Material Sciences".
The High Performance Computing Network of Northern Germany (HLRN) and computer facilities of the Freie Universit\"{a}t Berlin (ZEDAT) are acknowledged for computer time.
The authors are indebted to Filip Anselm Rasmussen and Kirsten Tr{\o}strup Winther (both Copenhagen) for providing the latest development version of the GPAW-\emph{GW} code as well as helpful discussions regarding its use. We would further like to express gratitude towards Lukas Hammerschmidt (Auckland) as well as Carmen Reden and Lisa Suntrup (both Berlin) for help with proofreading this manuscript.
The XCrySDen package  \cite{kokalj2003computer,kokalj1999xcrysden,kokalj2000scientific} was used to create images of atomic structures throughout this work while plots were created using Matplotlib \cite{Hunter_2007}.
\end{acknowledgments}

\bibliography{bibliography}{}

\end{document}